\documentclass[3p]{elsarticle}

\usepackage{amssymb}
\usepackage{amsmath}
\usepackage{lineno}
\usepackage{mathtools}

\journal{Applied Mathematics and Computation}

\begin{document}


\begin{frontmatter}

\title{Involution game with spatio-temporal heterogeneity of social resources}

\author[label1]{Chaoqian Wang}
\author[label2]{Attila Szolnoki}

\address[label1]{Program for Computational Social Science, Department of Computational and Data Sciences, George Mason University, Fairfax, VA 22030, USA}

\address[label2]{Institute of Technical Physics and Materials Science, Centre for Energy Research, P.O. Box 49, H-1525 Budapest, Hungary}

\begin{abstract}
When group members claim a portion of limited resources, it is tempting to invest more effort to get a larger share. However, if everyone acts similarly, they all get the same piece they would obtain without extra effort. This is the involution game dilemma that can be detected in several real-life situations. It is also a realistic assumption that resources are not uniform in space and time, which may influence the system's resulting involution level. We here introduce spatio-temporal heterogeneity of social resources and explore their consequences on involution. When spatial heterogeneity is applied, network reciprocity can mitigate the involution for rich resources, which would be critical otherwise in a homogeneous population. Interestingly, when the resource level is modest, spatial heterogeneity causes more intensive involution in a system where most cooperator agents, who want to keep investment at a low level, are present. This picture is partly the opposite in the extreme case when more investment is less effective. Spatial heterogeneity can also produce a counterintuitive effect when the presence of alternative resource levels cannot explain the emergence of involution. If we apply temporal heterogeneity additionally, then the impact of spatial heterogeneity practically vanishes, and we turn back to the behavior observed in a homogeneous population earlier. Our observations are also supported by solving the corresponding replicator equations numerically.
\end{abstract}


\end{frontmatter}

\section{Introduction}
\label{intro}

In a social dilemma, defective individuals gain more payoff at the expense of cooperative ones; meanwhile, the massive occurrence of defection leads to the lowest payoff for them, which is irrational for the group. How selfish individuals choose cooperation and how the rationality of the group prevails can be studied effectively by evolutionary dynamics  \cite{nowak_06,sigmund_10,duong_jmb19}. The evolutionary game theory principle assumes that individuals imitate strategies from a more successful partner. In structured populations where interactions are fairly fixed, this can be done by comparing payoff values with neighbors, resulting in cooperation supporting outcomes and fascinating images \cite{nowak1992evolutionary,perc2017statistical}. The simplest dilemma situations can be studied by pairwise interactions, such as the prisoner’s dilemma (PD), snowdrift (SD) and stag-hunt (SH) games \cite{da2015emergence,perc_bs10,duong_dga20}. By considering additional factors, such as incomplete information \cite{hilbe2018indirect} and multiple labels \cite{jensen2019discrimination,szolnoki_pre10}, we can further promote cooperation in pairwise interactions in the framework of the network reciprocity offered by structured populations \cite{ohtsuki2006simple}.

Naturally, we can also consider more subtle situations where the description based on pairwise interactions is insufficient, because the simultaneous presence of more partners may induce further effects that cannot be understood based on two-point interactions \cite{perc_jrsi13,burgio2020evolution,perc_srep15,battiston2020networks,szolnoki_epl16}. The most well-known and most intensively studied multiplayer game is the public goods game (PGG) which extends PD from pairwise interactions to group interactions \cite{yang_hx_pa19,li_k_csf21,chen_xj_srep16}. Similarly, additional factors have also been considered in multiplayer games to promote cooperation, such as reputation \cite{wei_x_epjb21,yang2019evolution,li2019reputation,wang2021promoting,quan2021reputation,ma2021effect,shen2022high}, punishment \cite{szolnoki_jtb13,wang2021tax,szolnoki_prx17,zhang2021conditional,lv2022particle,duong_prsa21,lee_hw_amc22}, exclusion \cite{liu_lj_srep17,zheng2021emergence,chen_xj_pcb18}, discounting and synergy 
\cite{quan2021reputation,quan2021effects}, fluctuating population size \cite{mcavoy2018public}, interdependence of different strategies \cite{szolnoki_pre15,wang2021public}, emerging alliance \cite{szolnoki_pre17,wang2021role}, environmental feedback \cite{szolnoki_epl17,yang2021environmental}, and reinvestment \cite{szolnoki2022tactical}. The possibility of real-world experiments \cite{tomassini2019computational} and applications to real scenarios was also an inspiring force along this research path.

Perhaps it is worth stressing that the types of multiplayer games are not limited to PGG, but also include N-person snowdrift games \cite{souza2009evolution}, N-person stag hunt games \cite{luo2021evolutionary}, N-person Hawk-Dove games \cite{chen2017evolutionary}, common-pool resource games \cite{lerat2013evolution}, and so on \cite{szolnoki_pre13}. In particular, a recently proposed multiplayer game is the so-called ``involution game,'' inspired by meaningless competition in a social group where available resources are fixed \cite{wang2021replicator,wang2022modeling}. Accordingly, the total payoff that all players can gain is constant, but each player receives a portion of the total payoff depending on the relative investments they made. In this way, the proportion of total payoff that each player receives depends on the ratio of the utility of their individual effort to the utility of all their collective efforts. As a natural reaction, a rational player is ready to invest more effort to gain a larger share, but if all think similarly, then the critical ratio remains intact; hence they obtain the same source that would be available reach for a lower price. In this way, keeping efforts low could be considered a cooperative act, while investing ``more'' is a sort of ``buying some privilege''; hence the latter is evaluated as a kind of defection.

Previous research introducing heterogeneity in evolutionary game models from different angles revealed a variety of new phenomena. One angle is the heterogeneity of networks, which was first raised by Santos and Pacheco~\cite{santos_prl05}, but recently Cimpeanu~{\it et al.}~\cite{cimpeanu2022artificial} also showed that network heterogeneity strongly influences safety development behaviours of advanced technologies. Another angle is the heterogeneity of game parameters. In pairwise games, for instance, heterogeneous parameters in the payoff matrix have been considered \cite{amaral2021criticality}. Similar approaches apply to multiplayer games. For the PGG, the heterogeneity of wealth \cite{wang2010effects}, allocation \cite{lei2010heterogeneity}, productivity (the synergy factor) \cite{perc2011does}, input \cite{ma2021effect,yuan2014role,huang2015effect,weng2021heterogeneous}, and the combinations of these have been studied \cite{zhang2017impact,fan2017promotion,liu2021effects}. In sum, it is generally believed that heterogeneity could be a cooperator-supporting condition, but it could also generate large inequality  \cite{mcavoy2020social}.

Although it is a fundamental assumption in the traditional involution game to have fixed social resources within each game group, this value should not necessarily be equal for all groups. Actually, the opposite scenario is more natural and can be observed in several real-life situations \cite{chen_xj_srep14}. For example, a worker in a developing country receives less payoff than another worker in a developed country devoting the same effort because their countries are endowed with different social resources. Or, a viewer who stands up in some cinemas will see more clearly (while blocking other viewers) than if he stands up in other cinemas because the quality of the screen varies from cinema to cinema. The social resources in different social entities are always not equal for participators to compete for. That is, we can assume heterogeneity of available resources for different groups. Of course, besides spatial heterogeneity, we may also consider temporal heterogeneity when the complete group benefit also changes in time \cite{szolnoki_srep19}.

Motivated by the above-described considerations, in this paper, we explore the possible consequences of heterogeneous resources in the involution game and reveal how they affect the fraction of defectors (i.e., the general involution level). A general framework of this type of work includes three complementary approaches: equilibrium calculations, evolutionary simulations, and behavioral experiments \cite{hilbe2019social}. This paper mainly focuses on evolutionary simulations in a structured population (Section 2 and 3). On the other hand, however, we also present equilibrium calculations in a well-mixed population as a complementary approach that helps understand the proper consequences of a structured population (Section 4). But first, we define our spatial model.

\section{Model}
\label{def}

We consider an involution game on an $L \times L$ square lattice with periodic boundary conditions where $N=L^2$ agents are distributed. Two strategies are available for an agent $i$ to choose during an elementary Monte Carlo (MC) step $t_{MC}$: investing less effort (cooperation, $S(i,t_{MC})=C$); or more (defection, $S(i,t_{MC})=D$) in the competition for social resources. Notably, every agents should bear a cost $o(i,t_{MC})$ depending on their choices. In particular, 
\begin{eqnarray}\label{investment}
o(i,t_{MC})=
\begin{cases} 
c,  & \mbox{if }S(i,t_{MC})=C,\\
d, & \mbox{if }S(i,t_{MC})=D,
\end{cases}
\end{eqnarray}
where $c<d$.

The cost's utility in the competition for resources, $u(i,t_{MC})$, which measures the competitiveness of agent $i$ at step $t_{MC}$, depends both on the cost and strategy. Namely, 
\begin{eqnarray}\label{utility}
u(i,t_{MC})=
\begin{cases} 
c,  & \mbox{if }S(i,t_{MC})=C,\\
\beta d, & \mbox{if }S(i,t_{MC})=D.
\end{cases}
\end{eqnarray}
Here $\beta>0$ is the relative utility of the defector strategy. If $\beta<1$ then more investment has less utility than less effort, and vice versa if $\beta>1$ \cite{wang2021replicator}. 

We denote the game group centered on agent $i$ by $\mathbb{N}(i)$, where the group size centered on agent $i$ is $|\mathbb{N}(i)|$. In this work, we use Von Neumann neighborhood where every agent interacts with their four nearest neighbors, hence $|\mathbb{N}(i)|=5$.

Similar to the PGG setup, an agent $i$ participates in $|\mathbb{N}(i)|$ involution games centered on itself and on each of its neighbors $j\in \mathbb{N}(i)$, averaging the payoff in each play as the final payoff $\pi(i,t_{MC})$. In the involution game centered around agent $j$, all $k\in \mathbb{N}(j)$ players involved in the group compete for social resources valued $\tilde{M}(j,t)$. This resource is divided among group members proportionally to each participant's competitiveness $u(k,t_{MC})$. In this way, the proportion of resources $\tilde{M}(j,t)$ that agent $i$ acquires from the specified competition is $u(i,t_{MC})/\sum_{k\in \mathbb{N}(j)}u(k,t_{MC})$. Evidently, the cost $o(i,t_{MC})$ should be subtracted. Therefore, the calculation of agent $i$'s payoff $\pi (i,t_{MC})$ at step $t_{MC}$ follows 
\begin{align}\nonumber
\pi(i,t_{MC})=&\frac{1}{|\mathbb{N}(i)|}\sum_{j\in \mathbb{N}(i)}\left(\frac{u(i,t_{MC})}{\sum_{k\in \mathbb{N}(j)}u(k,t_{MC})}
 \tilde{M}(j,t)-o(i,t_{MC})\right)\\
=&
\begin{cases} 
\displaystyle{\frac{1}{|\mathbb{N}(i)|}\sum_{j\in \mathbb{N}(i)}\left(\frac{c}{(n_C(j,t_{MC})+1)c+n_D(j,t_{MC})\beta d}
 \tilde{M}(j,t)-c\right)},  & \mbox{if }S(i,t_{MC})=C,\\
\displaystyle{\frac{1}{|\mathbb{N}(i)|}\sum_{j\in \mathbb{N}(i)}\left(\frac{\beta d}{n_C(j,t_{MC})c+(n_D(j,t_{MC})+1)\beta d}
 \tilde{M}(j,t)-d\right)}, & \mbox{if }S(i,t_{MC})=D,
\end{cases}\label{payoff}
\end{align}
where $n_C(j,t_{MC})$ and $n_D(j,t_{MC})$ respectively denote the numbers of cooperative and defective co-players (other than agent $i$) in the group centered around agent $j$ at step $t_{MC}$.

According to the evolutionary principle, we randomly select an agent $i$ and another agent $i'\in \mathbb{N}(i)$ to calculate their payoff values during an elementary step. Due to the pairwise comparison strategy updating protocol, player $i$ adopts the strategy of player $i'$ with the probability
\begin{eqnarray}\label{fermi}
p[S(i,t_{MC+1})\leftarrow S(i',t_{MC})]=\frac{1}{1+\mathrm{exp}\{[\pi(i,t_{MC})-\pi(i',t_{MC})]/\kappa \}}.
\end{eqnarray}
Here the parameter $\kappa>0$ characterizes the noise level of adoption \cite{szabo1998evolutionary}. Evidently, in the $\kappa \to 0$ limit, this update becomes fully rational, while for large $\kappa$ values, strategy change may also happen even if the payoff difference would not justify it. Notably, if we repeat the above described elementary step $N$ times, then we execute a full MC step when on average every player has a chance to update its strategy.

In the previous work we assumed a homogeneous system where every group used the same value of social resource $M$ \cite{wang2021replicator,wang2022modeling}. In our present work, however, we focus on a heterogeneous model, where this $M$ resource value may depend on space and time, rewritten by $\tilde{M}(i,t)$, indicating that the environment can be different for various groups and we are interested in whether it has any consequence on the resulting involution level. 

First, we consider a time-independent spatial heterogeneity of social resources by introducing a parameter $\eta_S$ ($\eta_S \ge0$). For each agent $i={1,\dots,N}$, we generate a random number $\chi(i)$ from a uniform distribution satisfying $-\eta_S \leq \chi(i) \leq \eta_S$. Then, we set 
\begin{eqnarray}\label{setm}
\tilde{M}(i,t)=M_0+\chi(i),
\end{eqnarray}
where the nonzero value of $\chi(i)$ leads to spatial heterogeneity and $\eta_S$ determines its degree. Evidently, $M_0$ represents the baseline of social resources, hence the average resource value 
\begin{eqnarray}\label{myields}
\int_{M_0-\eta_S}^{M_0+\eta_S}\tilde{M}(i,t)\,\mathrm{d}\tilde{M}(i,t)=\int_{-\eta_S}^{\eta_S}(M_0+\chi(i))\,\mathrm{d}\chi(i)=M_0
\end{eqnarray}
remains intact, no matter the specific value may change from place to place.

Second, we set a parameter $\eta_T$ ($0\le \eta_T\le 1$) to measure an additional temporal heterogeneity of social resources. At the very beginning of each full MC step $t$, we go through the whole population. For each $i={1,\dots,N}$ agent we regenerate a random number $\chi(i)$ with a probability $\eta_T$ and update $\tilde{M}(i,t)$ according to the new $\chi(i)$ value. Otherwise, $\tilde{M}(i,t)$ remains unchanged, hence $\tilde{M}(i,t)=\tilde{M}(i,t-1)$. After processing $\tilde{M}(i,t)$ for $i={1,\dots,N}$, we execute $N$ elementary MC steps in a usual way for a full MC step. Perhaps it is worth noting that in the suggested way spatial and temporal heterogeneities remain comparable.

To summarize the strategy updating procedure, we set system parameters $N$, $L$, $c$, $d$, $\beta$, $\kappa$, $M_0$, $\eta_S$, $\eta_T$, and initialize $S(i,t_{MC})$ and $\tilde{M}(i,t)$ for all $i={1,\dots,N}$ players (groups) before launching the simulation. In the starting state both strategies are distributed randomly among players with equal weights. At each full MC step $t$, we first go through $i={1,\dots,N}$ and update $\tilde{M}(i,t)$ by using the above described protocols. In this paper, we use $\kappa=0.1$, $c=1$, $L=400$, hence $N=160000$. But we note that qualitatively similar results can be obtained for other values. During the simulation we monitor the fraction of defector agents, denoted by $f_D$, which measures the average involution level. The stationary results of $f_D$ are obtained by running 10000 full MC steps and averaging its value over the last 2000 full MC steps. The key control parameters which determine the evolutionary outcome are $d$, $\beta$, $M_0$, $\eta_S$, and $\eta_T$. The main symbols we use in this work are listed in Table \ref{table1}.

\begin{table}[]
	\caption{Main symbols used in this work.}
	\label{table1}
	\centering
	\begin{tabular}{lll}
		\hline
		Symbol      & Interpretation                                                          & Property                                                                                  \\ \hline
		$L$         & The side length of the square.                                              & $L=400$.                                                                                  \\
		$N$         & The number of agents in the system.                                     & $N=L\times L=160000$.                                                                     \\
		$\kappa$    & The noise in selection.                                                 & $\kappa=0.1$.                                                                             \\
		$c$         & The cost of cooperation.                                                & $c=1$.                                                                                    \\
		$d$         & The cost of defection.                                                  & \begin{tabular}[c]{@{}l@{}}Independent parameter. \\ $d>c$.\end{tabular}                  \\
		$\beta$     & The relative utility of defection.                                      & Independent parameter.                                                                    \\
		$M_0$       & The baseline of social resources.                                       & Independent parameter.                                                                    \\
		$\eta_S$    & The spatial heterogeneity of social resources.                          & Independent parameter.                                                                    \\
		$\eta_T$    & The temporal heterogeneity of social resources.                         & Independent parameter.                                                                    \\
		$f_D^{(O)}$ & The defective fraction in a homogeneous population by MC simulations.   & Dependent variable.                                                                       \\
		$f_D^{(S)}$ & The defective fraction in a heterogeneous population by MC simulations. & Dependent variable.                                                                       \\
		$f_D^{(E)}$ & The expected fraction of defection in a heterogeneous population.       & \begin{tabular}[c]{@{}l@{}}Dependent variable, \\ calculated by $f_D^{(O)}$.\end{tabular} \\
		$f_D^{(C)}$ & The defective fraction by numerical replicator equations.               & Dependent variable.                                                                       \\ \hline
	\end{tabular}
\end{table}

\section{Results and discussion}
\label{spatial}

\begin{figure}[h!]
\centering
\includegraphics[width=15.0cm]{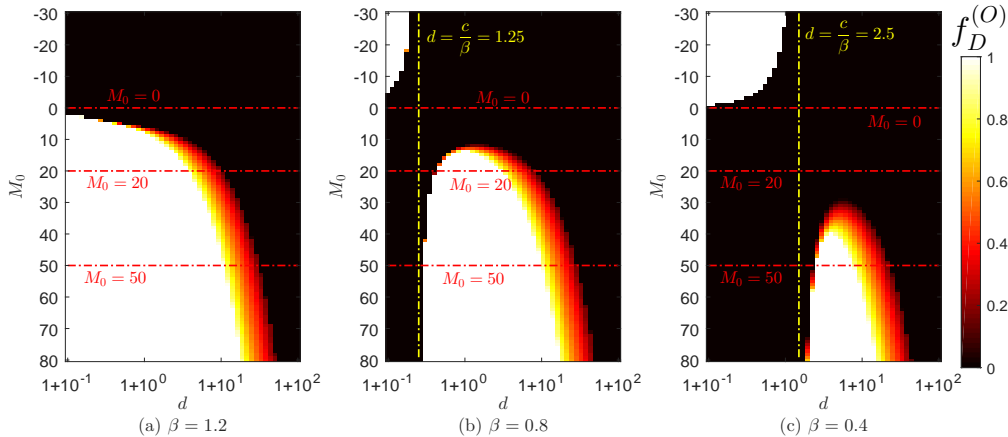}\\
\caption{The stationary fraction of defectors, $f_D^{(O)}$, as a bivariate function of control parameters $M_0$ and $d$ in the absence heterogeneities ($\eta_S =0$, $\eta_T = 0$). $M_0$ varies from $-30$ to $80$, with 111 data points for each $d$. $d$ varies from $1.1$ to $101$, with $50$ data points for each $M_0$. The panels show the system behavior for three representative values of relative utility parameter $\beta$. We can notice easily that the parameter $M_0$ is a decisive factor and the involution level depends sensitively on its value. For further reference, we marked the representative $M_0$ values which will serve as a basis for later cases when heterogeneity is applied. We also marked the critical $d$ values separating the full cooperator and full defector states in the large $M_0$ limit.}\label{md}
\end{figure}

As we already noted in the introduction, spatial populations may behave somewhat differently from the case when the interactions are randomized. Therefore, to gain a first impression, it is instructive to consider our spatial system in the absence of any heterogeneities. Technically it simply means that both $\eta_S$ and $\eta_T$ parameters, which characterize the degree of heterogeneities, are set to be zero. In this way, all groups benefit from the same level of resources. Fig.~\ref{md} summarizes the results of evolutionary simulations, where we plotted $f_D^{(O)}$, the stationary fraction of defectors. Note that superscript ``$O$" refers to the homogeneous, lack of resource heterogeneity case.

If we compare the system behavior to the well-mixed case presented in Ref.~\cite{wang2021replicator} then we can identify some generally valid features. For instance, in the case when more investment has more utility, shown in Fig.~\ref{md}(a), $f_D^{(O)}$ decreases monotonously as we increase $d$. For completeness, we can also simulate the region when $M_0<0$. In the area of $d<c/\beta$, strategy $D$ can prevail. This is because it receives fewer negative resources ($M_0<0$) although it devotes more effort than strategy $C$ does. Nevertheless, we do not give a realistic interpretation of the area of $M_0<0$, because the simulation of this area only intends to provide the original data points used for heterogeneous cases.

Qualitatively different behavior can be found for $\beta < 1$ when more investment has less utility. This situation is shown in Fig.~\ref{md}(b) and in Fig.~\ref{md}(c). If the resource is moderate, which means $M_0$ is not too high, then cooperators prevail for all $d$ values. For higher $M_0$ values, which means richer resources, however, there is a non-monotonous $d$-dependence of involution level. Here cooperators dominate for small $d<c/\beta$ values, which means it is better to keep involution at a low level. Beyond a critical $d$ value, however, the higher effort of defector strategy pays, and the system evolves to a full defector state. This solution remains valid until a critical $d$ value where the accompanying cost of defectors becomes intolerably high, hence cooperators prevail again. We note that qualitatively similar general behavior was also reported for the well-mixed case \cite{wang2021replicator}.

\begin{figure}[h!]
\centering
\includegraphics[width=15.0cm]{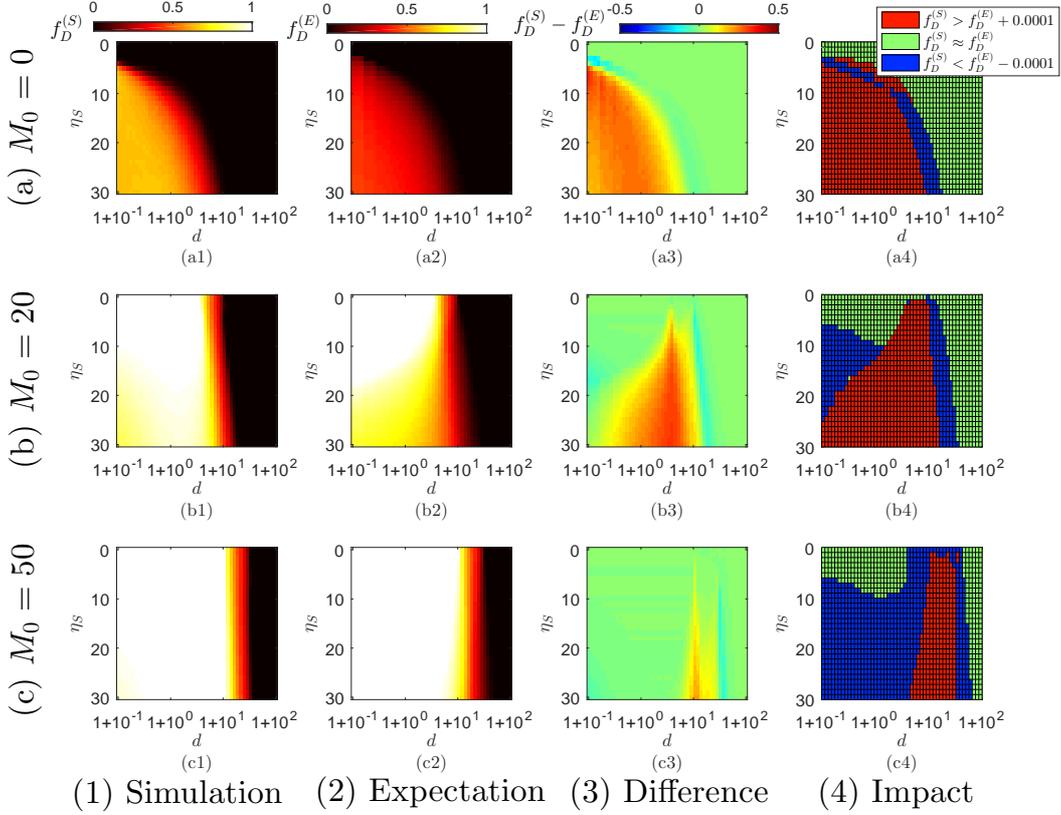}\\
\caption{The stationary fraction of defectors when spatial heterogeneity is introduced ($\eta_S \ge 0$), but temporal heterogeneity is not considered ($\eta_T=0$). In all panels, the control parameters are the ``relative" investment $d$ of defectors (because cooperators invest a unit) and $\eta_S$, the degree of spatial heterogeneity. Rows from (a) to (c) show cases at different general resource levels. These levels are marked by red lines in Fig.~\ref{md}. The first column shows the $f_D^{(S)}$ results of evolutionary simulations, while the second column, panels from (a2) to (c2), shows the $f_D^{(E)}$ expected fraction of defectors calculated by the weighted average due to different emerging resource levels according to Eq.~(\ref{datapro}). The third column depicts the $f_D^{(S)}-f_D^{(E)}$ difference between the proper and calculated values. The results shown in the last column can be considered a filtered or purified version of the previous column, which helps to reveal the proper impact of spatial heterogeneity. Accordingly, the red (blue) area denotes the parameter region where spatial heterogeneity provides higher (lower) general involution activity than the expected value obtained from the average of homogeneous cases at different source levels.In every cases $\beta=1.2$ was used.}\label{nasdbeta12}
\end{figure}

In the following, we introduce spatial heterogeneity of resources ($\eta_S \ge 0$), without considering temporal heterogeneity ($\eta_T=0$). First, we fix $\beta=1.2$, which represents the case when more investment has more utility. Our observations are summarized in Fig.~\ref{nasdbeta12}. Here we present the results obtained at three different cases where the average resource levels are $M_0=0$, $M_0=20$, and $M_0=50$, respectively. For comparison, these resource levels were already marked by red lines in Fig.~\ref{md}. The actual stationary fraction of defectors obtained from simulations is shown in the first column in Fig.~\ref{nasdbeta12}. Evidently, here the $\eta_S = 0$ border of the parameter plane marks what we obtain for a homogeneous system at the specific $M_0$ value. Starting from this line, if we fix the value of $d$ and increase $\eta_S$, then the involution level $f_D^{(S)}$ (``$S$" means the simulation results when introducing heterogeneities) may grow or decay depending on Fig.~\ref{nasdbeta12}(a1), Fig.~\ref{nasdbeta12}(b1), or Fig.~\ref{nasdbeta12}(c1) panel is considered. The lack of an obvious trend would be frustrating, but we can detect a certain rule in the data. In particular, if we start from a high involution level, which means $f_D^{(O)}$ is high, then $f_D^{(S)}$ becomes lower by increasing $\eta_S$. This trend can be seen in Fig.~\ref{nasdbeta12}(b1) and Fig.~\ref{nasdbeta12}(c1). However, the opposite is also true: if the involution level is low in the homogeneous case, then an increase in $\eta_S$ will elevate $f_D^{(S)}$. In other words, if we increase the degree of heterogeneity by increasing $\eta_S$, it will reverse the trend of involution level that we obtained for the related homogeneous system.

However, the proper consequence of spatial heterogeneity is more subtle because, as we illustrated in Fig.~\ref{md}, the involution level depends sensitively on the $M_0$ social resource level. For example, the increase of $M_0$ can elevate the involution level in a group significantly. However, there are also cases, depending on the original $M_0$ and $d$ values, when the change has no noticeable impact. Both scenarios can happen in a heterogeneous system where different groups should divide different values of the total payoff. To be more specific, in Fig.~\ref{nasdbeta12}(a1), where $M_0=0$, a positive $\eta_S$ involves $\tilde{M}(i,t)<0$ value for some groups. Referring to Fig.~\ref{md}(a), we can see that a negative resource level always indicates $f_D^{(O)}=0$ the full cooperation state. On the other hand, in a heterogeneous population, we also have $\tilde{M}(i,t)>0$ for other groups. Still referring Fig.~\ref{md}(a), we can see that positive $M_0$ value can easily result in an $f_D^{(O)}=1$ full defector state. Naturally, similar arguments can also be given for other $M_0$ values.

Therefore we can obtain a more realistic view about the consequence of spatially varying resource levels if we consider the whole set of available $\tilde{M}(i,t)$ values and calculate the resulting involution activity obtained for related homogeneous populations. The average of these values gives a better estimation of our expectations in a population where groups enjoy different resource levels. Therefore we introduce an estimated fraction of defectors, calculated as
\begin{eqnarray}\label{datapro}
f_D^{(E)}=\frac{1}{2\eta_S+1}\sum_{\tilde{M}=M_0-\eta_S}^{M_0+\eta_S}f_D^{(O)}.
\end{eqnarray}
We can see that Eq.~(\ref{datapro}) provides the average of $f_D^{(O)}$ values, shown in Fig.~\ref{md}, obtained for resource values in the range of $M_0-\eta_S\leq \tilde{M}(i,t)\leq M_0+\eta_S$. For instance, for $\beta=1.2$, $M_0=0$ and $\eta_S=10$ control parameter values we estimate $f_D^{(E)}$ from the average of the 21 original data points $f_D^{(O)}$ from $M_0=-10$ to $M_0=10$ in Fig.~\ref{md}(a). Similarly, we can compute $f_D^{(E)}$ for every $\eta_S-d$ pairs, shown in the second column of Fig.~\ref{nasdbeta12}.

We must stress that our calculation is just an approximation because how we averaged involution level ignores the fact that groups interact with each other. Indeed, the resulting plots in the second column are quite similar to the first column we obtained from actual MC simulations. However, still, they are not precisely equal. The difference between them indicates the real impact of spatial heterogeneity of social resources on the involution level. Therefore we calculate their difference, $f_D^{(S)}-f_D^{(E)}$, which is shown in the third column of Fig.~\ref{nasdbeta12}.

The resulting difference is rather complex, but we can generally interpret them in the following way. When $f_D^{(S)}-f_D^{(E)}>0$, then the actual fraction of cooperators exceeds the expected level indicating that spatial heterogeneity of social resources aggravates involution. In the opposite case, when $f_D^{(S)}-f_D^{(E)}<0$, we can say that spatial heterogeneity of resources inhibits involution. Last, when $f_D^{(S)}-f_D^{(E)}=0$, there is no significant interaction between neighboring groups hence the simulation results shown in the first column are just a simple superposition of the results obtained from groups enjoying different social resources.

Admittedly, it is hard to interpret the heat-map directly. Therefore we present the fourth column where we have divided the parameter plane into three regions and marked the cases mentioned above with different colors. In the first case, we plotted by red those parameter pairs where $f_D^{(S)}>f_D^{(E)}+0.0001$. Here we used a threshold value of 0.0001 because the accuracy of simulation results is limited; hence, this threshold helps us identify unambiguously those regions where the real involution level exceeds the estimated one. Similarly, we marked the parameter region by blue where $f_D^{(S)}<f_D^{(E)}-0.0001$, therefore spatiality truly helps to suppress involution activity. Last, we marked those parameter pairs by green where there is no significant difference between the measured and estimated value within the accuracy of simulations.

As expected, the fourth column is a real help in interpreting the consequence of spatial heterogeneity of resources. As Fig.~\ref{nasdbeta12}(a4) shows, the interaction of neighboring groups strengthens the involutions activity for the modest resource level, but this effect is not really ``dangerous'' because the $f_D^{(S)}$ is low; hence the involution is poor when $M_0=0$. In the opposite extreme case, when $M_0$ is high, shown in Fig.~\ref{nasdbeta12}(c4), the blue becomes significant, signaling that spatiality can mitigate the general involution. We stress that it is an important observation because previously, in a well-mixed system, we found that more abundant resources promote involution, which simply means that a rich environment could be harmful for the evolution of cooperation \cite{wang2021replicator}. However, this consequence is partly fixed in a spatially heterogeneous population where network reciprocity helps cooperators, as observed many times earlier.

\begin{figure}[h!]
\centering
\includegraphics[width=15.0cm]{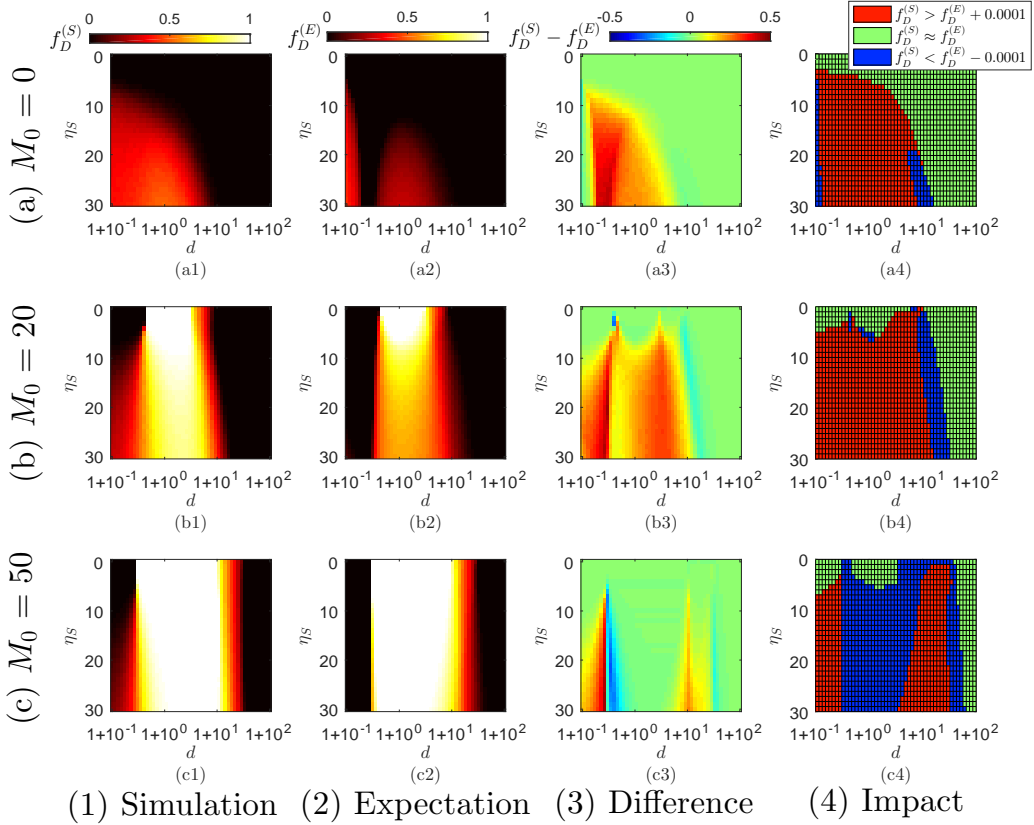}\\
\caption{The stationary fraction of defectors when pure spatial resource heterogeneity is applied. All parameters are identical to those we used in Fig.~\ref{nasdbeta12} except for $\beta=0.8$; hence the relative utility of more effort is less effective. In general, the main conclusions about the proper role of spatial resource heterogeneity are the same as we observed earlier for the $\beta>1$ case.}\label{nasdbeta08}
\end{figure}
In the following, we still apply pure spatial heterogeneity by keeping $\eta_T=0$ but consider $\beta=0.8$, which characterizes the situation when more investment is less effective. The results are summarized in Fig.~\ref{nasdbeta08} where we used the same setup we applied in Fig.~\ref{nasdbeta12}. Naturally, the heat maps are different from those we observed previously because not only the area where $d>c/\beta$ exists but also the area of $d<c/\beta$ emerges due to $\beta<1$. From the evolutionary simulations, shown in the first column, we can say that the access to various resources elevates involution; hence $f_D^{(S)}$ grows by increasing $\eta_S$. As previously, the expected $f_D^{(E)}$ values calculated by Eq.~(\ref{datapro}) is shown in the second column. Take Fig.~\ref{nasdbeta08}(a2) as an example, we can see that there is a pattern of $f_D^{(E)}>0$ and $f_D^{(E)}$ increases with $\eta_S$ in both the area of $d<c/\beta$ and $d>c/\beta$, while $f_D^{(E)}=0$ near $d=c/\beta$. It should be noted that the consequences of $f_D^{(E)}>0$ on the alternative sides of $d=c/\beta$ are different. Scanning up and down the line of $M_0=0$ in Fig.~\ref{md}(b), it is conceivable that the results for $f_D^{(E)}>0$ in the area of $d<c/\beta$ in Fig.~\ref{nasdbeta08}(a2) are originated from those local groups who have $\tilde{M}(i,t)<M_0$, while the results for $f_D^{(E)}>0$ in the area of $d>c/\beta$ come from those groups who enjoy $\tilde{M}(i,t)>M_0$ higher resource level. Furthermore, there is a qualitative difference between Fig.~\ref{nasdbeta08}(a1) and Fig.~\ref{nasdbeta08}(a2) because the mentioned area is connected in reality while it is separated into two parts in the estimated case.
\begin{figure}[h!]
\centering
\includegraphics[width=15.0cm]{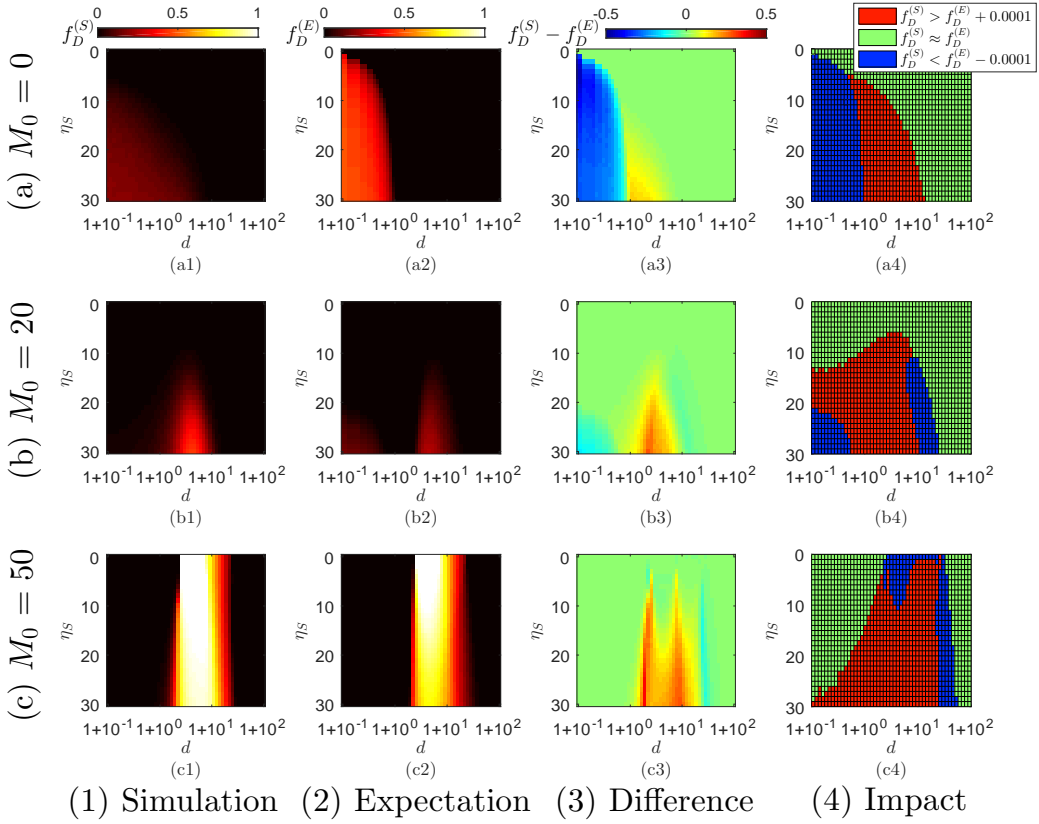}\\
\caption{
The stationary fraction of defectors when pure spatial resource heterogeneity is applied. All parameters are identical to those we used in Fig.~\ref{nasdbeta12} and Fig.~\ref{nasdbeta08}. The only difference is $\beta=0.4$ here. The last column highlights that the role of spatial heterogeneity of resources is conceptually different from those cases we discussed earlier. This is a straightforward consequence of a low $\beta$ value.}\label{nasdbeta04}
\end{figure}

Another interesting phenomenon can be detected in Fig.~\ref{nasdbeta08}(c1) where $M_0=50$ is used. In the $d<c/\beta$ area, we can observe positive $f_D^{(S)}$ value as $\eta_S$ increases, which is a kind of ``creating something from nothing'' effect. To give deeper insight, let us go back to the line of $M_0=50$ in Fig.~\ref{md}(b) and check its neighborhood. We can see that $f_D^{(O)}=0$ always holds in the $20\leq \tilde{M}(i,t)\leq 80$ region if $d$ is small. Therefore, given that $0\leq \eta_S\leq 30$, defectors have no chance to emerge in any local group. Still, $f_D^{(S)}$ becomes positive, as it is shown in Fig.~\ref{nasdbeta08}(c1), which should only be a spatial effect due to the interaction of networked groups. One may claim that positive $f_D^{(S)}$ can also be detected in Fig.~\ref{nasdbeta08}(b1) where $M_0=20$. In the latter case, however, some groups may have $\tilde{M}(i,t)\ge -10$ resource level, which can reach the full defector stage, as it is illustrated in Fig.~\ref{md}(b). Therefore a nonzero involution level can be expected for heterogeneous resources with $M_0=20$. However, not for $M_0=50$, as we explained, which merits the dramatic name we used above.

For a complete view, we have also repeated our simulations and made related calculations for the $\beta=0.4$ case, representing a situation when more investment is really ineffective. Our results are summarized in Fig.~\ref{nasdbeta04}. As in the above-discussed cases, the key finding can be found in the fourth column, where the proper role of spatial resource heterogeneity is evaluated. In the area of $d<c/\beta$, when $M_0$ is small, there is a parameter area where spatial heterogeneity tends to enhance cooperation (blue), which appears on the left side of Fig.~\ref{nasdbeta04}(a4). For larger $M_0$, this blue area shrinks and survives only for high $\eta_S$ values. It is also a generally valid observation that there is an optimal intermediate $d$ interval where significant resource heterogeneity helps to eliminate involution activity, and this effect is more pronounced for more prosperous resource conditions.
\begin{figure}[h!]
\centering
\includegraphics[width=15.0cm]{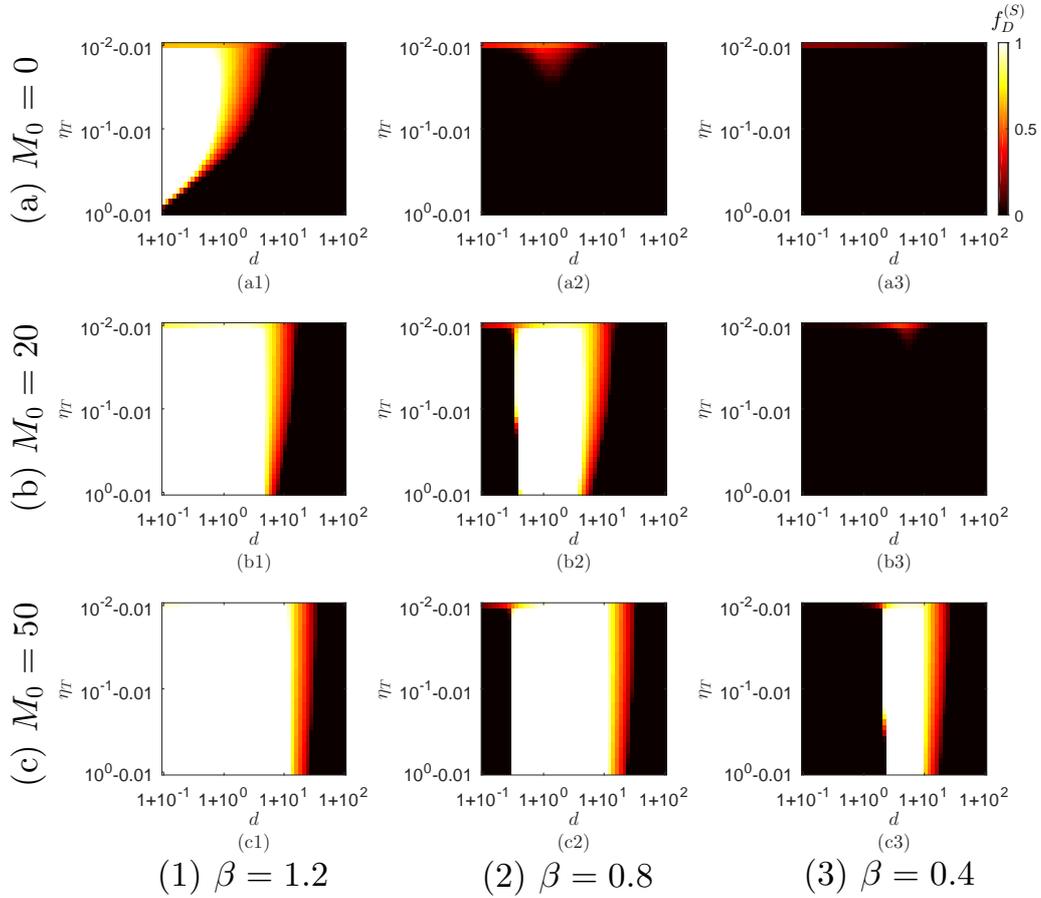}\\
\caption{The $f_D^{(S)}$ stationary fraction of defector strategy obtained from MC simulations when beside a permanent $\eta_S=30$ spatial heterogeneity we also imply temporal resource heterogeneity $\eta_T>0$ for groups. Rows from (a) to (c) show cases at different levels of average resources, as shown in the vertical axis. Each column shows results obtained at different $\beta$ values, representing various utility levels of more effort. The control parameters on heat maps are the range of temporal resource heterogeneity ($\eta_T$) and the $d$ relative investment of defector players. Note that we used logarithmic scales on both axes.}\label{natd}
\end{figure}

Interestingly, in our last case, the conclusion is somehow the opposite we reported earlier. More precisely, the spatial heterogeneity induced additional network reciprocity works more efficiently when the average resource level $M_0$ is low and more efficient for a rich resource case. This observation is essentially expected because low $\beta$ makes defectors weak in general hence heterogeneous resources cannot add anything relevant to the system behavior. Therefore we here basically witness conceptually similar behavior we reported for a homogeneous population earlier.

Next, besides spatial, we also add temporal heterogeneities to social resources to explore their collective impacts on the involution activity. Practically it means that both $\eta_S$ and $\eta_T$ parameters are positive. It is easy to see that the smaller the applied spatial heterogeneity $\eta_S$, the more limited the reachable impact of temporal heterogeneity $\eta_T$ on the results. It is because the practical consequence of nonzero $\eta_S$ is to provide a $[M_0-\eta_S, M_0+\eta_S]$ range of resources for each local group. Evidently, this effect is marginal for small spatial heterogeneities. Hence, in the $\eta_S=0$ limit, nonzero $\eta_T$ usage has no detectable impact on the results. Therefore, to observe the possible effect more easily, we apply $\eta_S=30$, which is the maximum degree of permanent spatial heterogeneity range in Fig.~\ref{nasdbeta12}, Fig.~\ref{nasdbeta08}, and Fig.~\ref{nasdbeta04}. As the mentioned figures illustrate, this degree value can ensure the maximal consequence of spatial heterogeneity. 

The results of our MC simulation are summarized in Fig.~\ref{natd} where we used $\eta_T$ and $d$ as control parameters to show the $f_D^{(S)}$ fraction of cooperator strategy. Horizontally we show results obtained at different $M_0$ average source values, while vertically, we used alternative $\beta$ utility levels of more effort used by defector strategy. For easier comparison, we apply the same parameter values we used in earlier figures in both cases. We note that both axes of heat maps are logarithmic here. Evidently, when $\eta_T=0$, we get back the results shown in the first column in previous figures at $\eta_S=30$ horizontal line. As we apply more intensive temporal heterogeneity by increasing $\eta_T$, the way how $f_D^{(S)}$ changes depends slightly on other parameters. However, there is a general valid observation. Namely, when $\eta_T\gg 0$, the resulting involution levels resemble those we obtained at $\eta_S=0$ in previous figures.

This observation actually further supports our previous findings of how spatial heterogeneity of resources modifies the general involution level. As we argued, the main impact is based on the fact that groups may enjoy different resource levels, which determines the emerging pattern of strategies. Once this fixed structure is broken by applying $\eta_T>0$, the consequence of the external resource condition cannot be maintained anymore. As a result, the impact of spatial heterogeneities practically vanishes, and we turn back to the behavior observed in a homogeneous population. Fig.~\ref{natd} also illustrates that this effect occurs relatively early, no matter $\eta_T$ values are shown on a logarithmic scale.

\section{Equilibrium calculations in an infinite well-mixed population}
\label{equilibrium}

To complete our study, in the remainder of this work, we present mean-field calculations of a model which assumes an infinitely large population. For this goal, we extend and generalize the method of replicator dynamics used in Ref.~\cite{wang2021replicator}. We stress that this section is just a complementary approach using different dynamics, but it can help identify the generally valid behaviors. 

Since the population is infinite, the individuals in each case enjoying $\tilde{M}$ resource level are also infinite. Also, there are infinite subsets of individuals who can be distinguished by the available $\tilde{M}$ value. Accordingly, in analogy with Eq.~(\ref{payoff}), a player's $\pi(\tilde{M})$ payoff who reaches $\tilde{M}$ is the following:
\begin{align}\nonumber
\pi(\tilde{M})=&\frac{u}{n_Cc+n_D\beta d+u} \tilde{M}-o\\
=&
\begin{cases} 
\displaystyle{\frac{c}{(n_C+1)c+n_D\beta d} \tilde{M}-c\coloneqq \pi_C(\tilde{M})},  & \mbox{if }S=C,\\
\displaystyle{\frac{\beta d}{n_Cc+(n_D+1)\beta d} \tilde{M}-d\coloneqq \pi_D(\tilde{M})}, & \mbox{if }S=D.
\end{cases}\label{singlepayoff}
\end{align}

Next, because of the limited feasibility of the applied mathematical technique, we consider two special cases, which are $\eta_T=0$ and $\eta_T\gg 0$.

\subsection{The case for $\eta_T=0$}
\label{T0}

As a reference, we first study the $\eta_T=0$ case when each player has access to a permanent resource value. The average payoff for cooperators and for defectors who reach $\tilde{M}$ resource level is
\begin{subequations}\label{avepayoff}
\begin{align}
\langle \pi_C(\tilde{M})\rangle =\sum_{n_D=0}^{n-1}\dbinom{n-1}{n_D}\langle y\rangle ^{n_D}(1-\langle y\rangle )^{n-n_D-1} \pi_C(\tilde{M}),\tag{\ref{avepayoff}{a}} \label{avepic}\\
\langle \pi_D(\tilde{M})\rangle =\sum_{n_D=0}^{n-1}\dbinom{n-1}{n_D}\langle y\rangle ^{n_D}(1-\langle y\rangle )^{n-n_D-1} \pi_D(\tilde{M}).\tag{\ref{avepayoff}{b}} \label{avepid}
\end{align}
\end{subequations}
where $n$ denotes the number of players in a group, hence $n_C+n_D=n-1$, and $\langle y\rangle$ is the average fraction of defectors in the whole population. The focal individual who is directly linked to $\tilde{M}$ resource can play involution games with others from all categories who also participate in various $\tilde{M}$. Since the distribution of $\tilde{M}$ is uniform, $\langle y\rangle$ should be the average fraction of defectors from all categories with $\tilde{M}\in [M_0-\eta_S,M_0+\eta_S]$, and is calculated as  
\begin{eqnarray}\label{aveysatis}
\langle y\rangle=\frac{\int_{M_0-\eta_S}^{M_0+\eta_S}y(\tilde{M})\mathrm{d}\tilde{M}}{\int_{M_0-\eta_S}^{M_0+\eta_S}\mathrm{d}\tilde{M}}=\frac{1}{2\eta_S}\int_{M_0-\eta_S}^{M_0+\eta_S}y(\tilde{M})\mathrm{d}\tilde{M},
\end{eqnarray}
where $y(\tilde{M})$ is the fraction of defectors who are linked to $\tilde{M}$ resource value.

By choosing a simple, but consistent evolutionary description, we apply the Fermi updating rule in replicator dynamics \cite{wang2020interplay}. For each subset with $\tilde{M}$, the updating of strategies following its replicator dynamics, which can be written as
\begin{eqnarray}\label{ymdynamics}
\dot y(\tilde{M})=y(\tilde{M})(1-y(\tilde{M}))\dfrac{1}{1+\mathrm{e}^{-\dfrac{\langle \pi_D(\tilde{M})\rangle-\langle \pi_C(\tilde{M})\rangle}{\kappa}}}-(1-y(\tilde{M}))y(\tilde{M})\dfrac{1}{1+\mathrm{e}^{-\dfrac{\langle \pi_C(\tilde{M})\rangle-\langle \pi_D(\tilde{M})\rangle}{\kappa}}}.
\end{eqnarray}
In Eq.~(\ref{ymdynamics}), strategies are only contagious among individuals belonging to the same subset (distinguished by $\tilde{M}$), which is a standard approach in replicator dynamics \cite{wang2010effects,wang2020interplay}.

For the whole system, the time evolution of the fraction of defectors, $\dot{\langle y\rangle}$ (different from $\langle \dot{y}\rangle$), is depicted by using the derivative of $\langle y\rangle$  with respect to $t$: 
\begin{eqnarray}\label{aveydynamics}
\dot{\langle y\rangle}=\frac{\mathrm{d}}{\mathrm{d}t}\langle y\rangle=\frac{1}{2\eta_S}\int_{M_0-\eta_S}^{M_0+\eta_S}\dot y(\tilde{M})\mathrm{d}\tilde{M}.
\end{eqnarray}

We note that the stability of $y(\tilde{M})$ for $\forall \tilde{M}\in [M_0-\eta_S,M_0+\eta_S]$ is not a necessary condition of the equilibrium for $\langle y\rangle$. For example, $\dot{\langle y\rangle}=0$ may hold when $\dot y(\tilde{M})<0$ for some $\tilde{M}$ and $\dot y(\tilde{M})>0$ for others. However, we here intuitively define $\langle y\rangle$ achieving stability by $y(\tilde{M})$ achieving stability for $\forall \tilde{M}\in [M_0-\eta_S,M_0+\eta_S]$. Based on this definition and the domain $0\leq y^*(\tilde{M})\leq 1$, the system of Eq.~(\ref{aveydynamics}) has at least two equilibria: (i) $\langle y\rangle ^*=0$, cooperation exists only, which holds if $y^*(\tilde{M})=0$ for $\forall \tilde{M}\in [M_0-\eta_S,M_0+\eta_S]$; (ii) $\langle y\rangle ^*=1$, defection exists only, which holds if $y^*(\tilde{M})=1$ for $\forall \tilde{M}\in [M_0-\eta_S,M_0+\eta_S]$. There may also be other equilibria in $0<\langle y\rangle ^*<1$, where cooperation and defection coexist. As reported previously \cite{wang2021replicator}, it is challenging to study the inner equilibrium point, where $0<\langle y\rangle ^*<1$, directly. But we can overcome this difficulty because the stability of $\langle y\rangle ^*=0$ and $\langle y\rangle ^*=1$ extreme points is feasible. If both are unstable, then we can conclude that cooperation and defection coexist.

For $\forall \tilde{M}\in [M_0-\eta_S,M_0+\eta_S]$, the first-order Jacobian matrix of Eq.~(\ref{ymdynamics}) is 
\begin{align}\nonumber
\frac{\mathrm{d}\dot y(\tilde{M})}{\mathrm{d}y(\tilde{M})}=
&\left[\frac{\mathrm{d}}{\mathrm{d}y(\tilde{M})}y(\tilde{M})(1-y(\tilde{M}))\right]
\left(\dfrac{1}{1+\mathrm{e}^{-\dfrac{\langle \pi_D(\tilde{M})\rangle-\langle \pi_C(\tilde{M})\rangle}{\kappa}}}-\dfrac{1}{1+\mathrm{e}^{-\dfrac{\langle \pi_C(\tilde{M})\rangle-\langle \pi_D(\tilde{M})\rangle}{\kappa}}}\right)\\
&+y(\tilde{M})(1-y(\tilde{M}))
\left[\frac{\mathrm{d}}{\mathrm{d}y(\tilde{M})}\left(\dfrac{1}{1+\mathrm{e}^{-\dfrac{\langle \pi_D(\tilde{M})\rangle-\langle \pi_C(\tilde{M})\rangle}{\kappa}}}-\dfrac{1}{1+\mathrm{e}^{-\dfrac{\langle \pi_C(\tilde{M})\rangle-\langle \pi_D(\tilde{M})\rangle}{\kappa}}}\right)\right],\label{jacobian}
\end{align}
where, according to Eq.~(\ref{avepayoff}), we have
\begin{eqnarray}\label{avepaygap}
\langle \pi_D(\tilde{M})\rangle -\langle \pi_C(\tilde{M})\rangle =\sum_{n_D=0}^{n-1}\dbinom{n-1}{n_D}\langle y\rangle ^{n_D}(1-\langle y\rangle )^{n-n_D-1}(\pi_D(\tilde{M}) -\pi_C(\tilde{M})).
\end{eqnarray}

First, we study the stability of $\langle y\rangle ^*=0$, which is equivalent to the stability of $y^*(\tilde{M})=0$ for $\forall \tilde{M}\in [M_0-\eta_S,M_0+\eta_S]$. In this case, the only non-zero term of Eq.~(\ref{avepaygap}) is the term of $n_D=0$, which results in 
$\langle \pi_D(\tilde{M})\rangle -\langle \pi_C (\tilde{M})\rangle =\pi_D (\tilde{M})-\pi_C (\tilde{M})$. Also, only the first line of Eq.~(\ref{jacobian}) is non-zero. Therefore, Eq.~(\ref{jacobian}) is given by 
\begin{eqnarray}\label{jacoy0}
\left . \frac{\mathrm{d}\dot y(\tilde{M})}{\mathrm{d}y(\tilde{M})}\right|_{y(\tilde{M})=0}=\left .\left(\dfrac{1}{1+\mathrm{e}^{-\dfrac{\pi_D(\tilde{M})- \pi_C(\tilde{M})}{\kappa}}}-\dfrac{1}{1+\mathrm{e}^{-\dfrac{ \pi_C(\tilde{M})- \pi_D(\tilde{M})}{\kappa}}}\right)\right|_{y(\tilde{M})=0}.
\end{eqnarray}

According to a basic calculus, $y^* (\tilde{M})=0$ is stable if $\mathrm{d}\dot y(\tilde{M})/\mathrm{d}y(\tilde{M})|_{y(\tilde{M})=0}<0$, which is equivalent to $\pi_D(\tilde{M})-\pi_C(\tilde{M})<0$ due to Eq.~(\ref{jacoy0}). Therefore, $\langle y\rangle ^*=0$ is stable if $\pi_D(\tilde{M})-\pi_C(\tilde{M})<0$ for $\forall \tilde{M}\in [M_0-\eta_S,M_0+\eta_S]$. By using Eq.~(\ref{singlepayoff}), $\pi_D (\tilde{M})-\pi_C(\tilde{M})<0$ can be calculated as
\begin{eqnarray}\label{condi0}
\left(\frac{\beta d}{(n-1)c+\beta d}-\frac{1}{n}\right) \tilde{M}-(d-c)<0.
\end{eqnarray}
The left side of Eq.~(\ref{condi0}) increases with $\tilde{M}$ if  $[c/(\beta d)-1](n-1)<0$, and vice versa. In multiplayer games, $n>1$, hence the condition is equivalent to $c/(\beta d)-1<0$. Therefore, when $d>c/\beta$, the left side of Eq.~(\ref{condi0}) increases with $\tilde{M}$. Since $\tilde{M}\leq M_0+\eta_S$, the sufficient condition of stability is 
\begin{eqnarray}\label{Q1-}
\mathcal Q_1^-\eqqcolon \left (\frac{\beta d}{(n-1)c+\beta d}-\frac{1}{n}\right)(M_0+\eta_S)-(d-c)<0.
\end{eqnarray}
Similarly, when $d<c/\beta$, the left side of Eq.~(\ref{condi0}) decreases with $\tilde{M}$. Since $\tilde{M}\ge M_0-\eta_S$, the sufficient condition of stability is
\begin{eqnarray}\label{Q2-}
\mathcal Q_2^-\eqqcolon \left (\frac{\beta d}{(n-1)c+\beta d}-\frac{1}{n}\right)(M_0-\eta_S)-(d-c)<0.
\end{eqnarray}

In sum, we obtain two curves $\mathcal Q_1^-=0$ and $\mathcal Q_2^-=0$ in parameter space $d>c/\beta$ and $d<c/\beta$, respectively, indicating the transition points between the full cooperation and coexisting phases.

Secondly, we study the stability of $\langle y\rangle ^*=1$. The only non-zero part of Eq.~(\ref{avepaygap}) is the term of $n_D=n-1$, which also reduces Eq.~(\ref{avepaygap}) to $\langle \pi_D(\tilde{M})\rangle -\langle \pi_C(\tilde{M})\rangle =\pi_D (\tilde{M})-\pi_C (\tilde{M})$. Then, Eq.~(\ref{jacobian}) can be calculated as
\begin{eqnarray}\label{jacoy1}
\left . \frac{\mathrm{d}\dot y(\tilde{M})}{\mathrm{d}y(\tilde{M})}\right|_{y(\tilde{M})=1}=\left .-\left(\dfrac{1}{1+\mathrm{e}^{-\dfrac{\pi_D(\tilde{M})- \pi_C(\tilde{M})}{\kappa}}}-\dfrac{1}{1+\mathrm{e}^{-\dfrac{ \pi_C(\tilde{M})- \pi_D(\tilde{M})}{\kappa}}}\right)\right|_{y(\tilde{M})=1}.
\end{eqnarray}
Here the condition $\mathrm{d}\dot y(\tilde{M})/\mathrm{d}y(\tilde{M})|_{y(\tilde{M})=1}<0$ requires $-(\pi_D(\tilde{M})-\pi_C(\tilde{M}))<0$. Therefore $\langle y\rangle ^*=1$ is stable if $-(\pi_D(\tilde{M})-\pi_C(\tilde{M}))<0$ for $\forall \tilde{M}\in [M_0-\eta_S,M_0+\eta_S]$, which is
\begin{eqnarray}\label{condi1}
\left(\frac{c}{(n-1)\beta d+c}-\frac{1}{n}\right) \tilde{M}+(d-c)<0.
\end{eqnarray}
When $d>c/\beta$, the sufficient condition of stability is
\begin{eqnarray}\label{Q1+}
\mathcal Q_1^+\eqqcolon \left(\frac{c}{(n-1)\beta d+c}-\frac{1}{n}\right)(M_0-\eta_S)+(d-c)<0.
\end{eqnarray}
For $d<c/\beta$, the sufficient condition of stable equilibrium is
\begin{eqnarray}\label{Q2+}
\mathcal Q_2^+\eqqcolon \left(\frac{c}{(n-1)\beta d+c}-\frac{1}{n}\right)(M_0+\eta_S)+(d-c)<0.
\end{eqnarray}
We now have $\mathcal Q_1^+=0$ and $\mathcal Q_2^+=0$ curves in the parameter space $d>c/\beta$ and $d<c/\beta$, respectively, separating the full defector and coexistence phases.

\begin{figure}
\centering
\includegraphics[width=15.0cm]{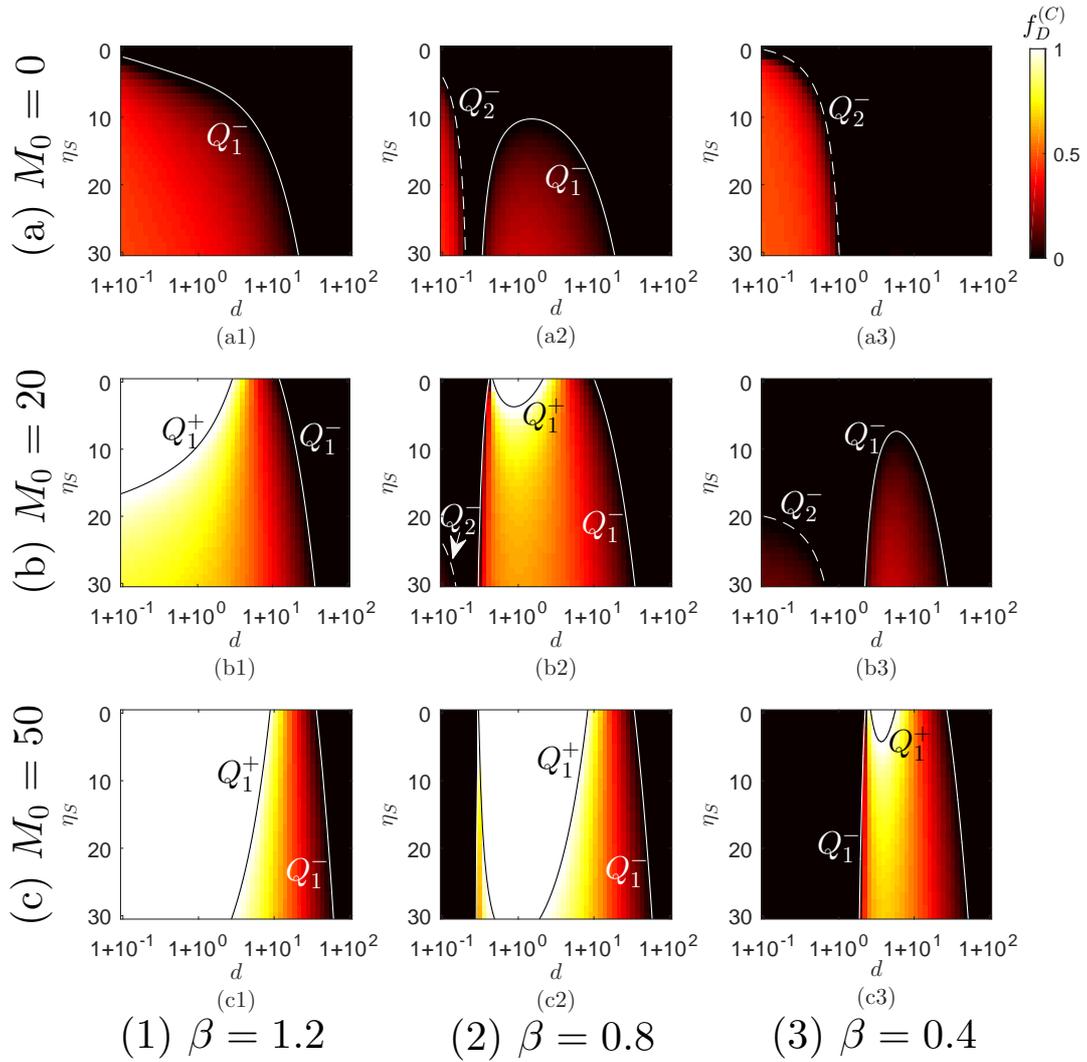}\\
\caption{The $f_D^{(C)}$ stationary fraction of defector strategy obtained form the numerical calculation of Eq.~(\ref{aveydynamics}) when $n=5$. Rows from (a) to (c) show cases at different levels of average resources, as shown in the vertical axis. Each column shows results obtained at different $\beta$ values; hence they represent various utility levels of more effort. The control parameters on heat maps are the range of spatial resource heterogeneity ($\eta_S$) and defector players' relative investment $d$. The analytical curves, which are defined in the main text, mark the borderlines of different phases. Here we have full cooperator, full defector phases, and between them, a mixed phase where strategies coexist.}\label{analyfig}
\end{figure}

The results of our numerical calculations are summarized in  Fig.~\ref{analyfig} where we plotted the $f_D^{(C)}$ equilibrium fraction of defectors. Similar to previous figures, rows show results obtained at different levels of average resources. Here we use the same $M_0$ values applied earlier. Columns depict cases obtained at different utility levels of defector investments. Again, these $\beta$ values are equal to those we used in other figures for proper comparison. The control parameters are the range of spatial heterogeneity $\eta_S$ and enhanced investment of defectors, $d$.

The color code which characterizes the involution level is from numerical calculation of Eq.~(\ref{aveydynamics}) when $\dot {\langle y\rangle}=0$. The lines are the solutions of  $\mathcal Q_1^-=0$, $\mathcal Q_2^-=0$ and $\mathcal Q_1^+=0$, where $\mathcal Q_1^-$, $\mathcal Q_2^-$ and $\mathcal Q_1^+$ are defined by Eq.~(\ref{Q1-}), Eq.~(\ref{Q2-}) and Eq.~(\ref{Q1+}). They mark the borderlines separating the full cooperator, full defector, and coexistence phases. Note that the curve of $\mathcal Q_2^+=0$ is invisible because it always hides in the area of $M_0<0$ according to Eq.~(\ref{Q2+}). In addition, it is worth mentioning that Fig.~\ref{analyfig} resembles the expectation results in second column of Figs.~(\ref{nasdbeta12}$-$\ref{nasdbeta04}) qualitatively. This is because both approaches assumed the unstructured population condition.

\subsection{The case for $\eta_T\gg 0$}
\label{T}

The other case which can be calculated easily is the strong temporal heterogeneity ($\eta_T\gg 0$) limit. Here the subset of resources where an individual belongs is the statistical average of all available $\tilde{M}\in [M_0-\eta_S,M_0+\eta_S]$ values. Accordingly, the expected payoff for a cooperator player is
\begin{subequations}\label{avepayoff2}
\begin{align}
\langle \pi_C(\tilde{M})\rangle
=&\int_{M_0-\eta_S}^{M_0+\eta_S}\sum_{n_D=0}^{n-1}\dbinom{n-1}{n_D}\langle y\rangle ^{n_D}(1-\langle y\rangle )^{n-n_D-1} \pi_C(\tilde{M})\mathrm{d}\tilde{M}\notag \\
=&\sum_{n_D=0}^{n-1}\dbinom{n-1}{n_D}\langle y\rangle ^{n_D}(1-\langle y\rangle )^{n-n_D-1} \pi_C(M_0).\tag{\ref{avepayoff2}{a}} \label{avepic2}
\end{align}
\end{subequations}
Similarly, the expected defector's income is
\begin{subequations}\nonumber
\begin{align}
\langle \pi_D(\tilde{M})\rangle =\sum_{n_D=0}^{n-1}\dbinom{n-1}{n_D}\langle y\rangle ^{n_D}(1-\langle y\rangle )^{n-n_D-1} \pi_D(M_0).\tag{\ref{avepayoff2}{b}}\label{avepid2}
\end{align}
\end{subequations}
If we check these expressions, we can see that these values agree with those we obtained for the classical model where all groups obtain $M_0$ resource. Consequently, the conclusion is similar, hence the strong temporal heterogeneity limit $\eta_T\gg 0$ gives back the system behavior we observed in the homogeneous, $\eta_T=0$ and $\eta_S=0$ classical model.

\section{Conclusion}

In social dilemmas, cooperation always assumes a coordinated action from competitors while a defector generally enjoys behaving differently from the others. This is exactly the case in a situation described by the involution game. When the reachable resource divided among group members is limited, some players may want to invest more to get a large pie from the common pizza. Others should also invest more to restore the original fragments and avoid being overtaken. As a result, everybody gets the original share, but for a higher price. Accordingly, in the above-described situation, those behave as cooperators who coordinate their acts and try the necessary investment at a low level. A seemingly paradoxically, those who invest more are the defectors because they generate a larger investment activity. Our previous work obtained in a well-mixed population has pointed out that the fraction of defectors, which means the general degree of involution, depends sensitively on the actual resource value of the group. However, this level should not necessarily be equal for all groups. This can be justified by varying environments or other time-dependent factors. Motivated by this observation in the present work, we have studied the possible consequences of spatio-temporal heterogeneity in the available social resources.

First, we have considered the case of fixed but spatially heterogeneous resource distribution. Our Monte Carlo simulation highlighted that the proper interaction between neighboring groups could mitigate the involution level when the resource value is generally high. This is a positive consequence of network reciprocity because, in the absence of resource heterogeneity, the involution level would be high in these circumstances. The effect of spatial heterogeneity, however, is not as straightforward because in a poor resource environment, when involution would be low in a homogeneous system, the heterogeneity induces a slight growth of involution. Luckily, the resulting global involution level is still tolerable.

For completeness, we also have studied the case when the utility of additional investment of defectors is less effective. Let us stress that it is mostly just a theoretical option because, in this case, defectors are not strongly motivated to invest more. Interestingly, however, spatial heterogeneity supports the original effect we observed in homogeneous populations in this partly exotic case.

Furthermore, we have also added temporal heterogeneities to spatially varying resource distribution. Here the key finding is that the additional temporal heterogeneity practically vanishes the impact of spatial heterogeneity. The final system behavior is familiar to the one we observed in a homogeneous population at the given resource level. We have also solved the related replicator equation numerically and found conceptually similar system behavior to support simulation observations.

It is a frequently observed and broadly accepted observation that heterogeneities can elevate the cooperation level in a system formed by competitors with conflicting individual interests \cite{szolnoki_epl07,perc_pre08,santos_n08,chen_xj_njp14}. The general argument explains it because heterogeneity involves largely different payoff distribution, which helps more successful players coordinate their neighborhoods. However, this process helps cooperator strategy to gain higher success and is harmful to defectors who cannot exploit neighbors anymore. In our present study, a whole group needed to share a certain resource value; therefore, the interaction of groups becomes more important. Our key conclusion is that heterogeneity may have positive consequences, but the picture is more subtle when the external environment is less supportive for defectors. 

\bibliographystyle{elsarticle-num-names}

\end{document}